\documentclass{llncs}
\usepackage{graphicx}
\usepackage{colortbl}
\usepackage{subfigure}
\usepackage{multirow}
\usepackage{url}
\usepackage{algorithmic}

\begin{document}
\pagestyle{empty}
\mainmatter
\title{An Evaluation of Link Neighborhood Lexical Signatures to Rediscover Missing Web Pages}
\author{Jeb Ware \and Martin Klein \and Michael L. Nelson}

\institute{Old Dominion University, Department of Computer Science\\
Norfolk VA 23529}

\maketitle

\begin{center}
\{jware, mklein, mln\}@cs.odu.edu
\end{center}
\begin{abstract}
For discovering the new URI of a missing web page, lexical signatures, which consist of a small number of words chosen to represent the ``aboutness'' of a page, have been previously proposed.
However, prior methods relied on computing the lexical signature before the page was lost, or using cached or archived versions of the page to calculate a lexical signature.  
We demonstrate a system of constructing a lexical signature for a page from its link neighborhood, that is the ``backlinks'', or pages that link to the missing page.
After testing various methods, we show that one can construct a lexical signature for a missing web page using only ten backlink pages.
Further, we show that only the first level of backlinks are useful in this effort.
The text that the backlinks use to point to the missing page is used as input for the creation of a four-word lexical signature.
That lexical signature is shown to successfully find the target URI in over half of the test cases.
\end{abstract}
\section{Introduction} \label{sec:intro}
At times during a web browsing session, a user's navigation is interrupted
by broken links which point to non-existent pages.  In some cases, these
pages are not lost forever, but can be accessed at a different address.
Lexical signatures, which consist of a small number of words chosen to
represent the ``aboutness'' of a page, have been proposed as a way to
discover the new address of a page.  For example, a lexical signature
for a page that sells instruments for environmental measurement
might be [humidity, measurement, machine].  Prior research has shown
\cite{opal:ht06,klein:revisiting,1028101} that using those words as
a query to a search engine may yield the same page at another URI.
However, previous methods relied on computing the lexical signature
before the page was lost, or using cached or archived versions of the
page to calculate a lexical signature.  If the lexical signature had not
been previously calculated, and no archived or cached copies of the page
were available, these methods are unusable.

To overcome this limitation, an experiment was performed to evaluate
constructing lexical signatures from link neighborhoods.  Since pages tend
to link to related pages, the intuition was that the link neighborhoods
contain enough of the ``aboutness'' of the targetted page to allow a
lexical signature to be created.  Link neighborhoods were constructed by
querying a search engine for listings of backlinks, and several methods
of calculating lexical signatures from those link neighborhoods were
tested to see which method produced the most effective signatures.

The effects of lexical signature size, backlink depth, and backlink ranking were examined, as well as the radius within a backlink page from which terms for the lexical signature should be drawn.
The method most likely to succeed used three terms in the lexical signature, the terms being drawn only from the anchor text of links pointing to the missing page.
It was necessary to use only the first ten backlinks for a URI to calculate the lexical signature.
Further, only first-level backlinks, those pages that linked directly to the missing page, proved useful; second-level backlinks yielded poorer results.
%
%
\section{Lexical Signatures} \label{sec:lex-sigs}
\label{sec:lex-sigs}
\begin{figure}[t!]
 \centering
 \includegraphics[scale=0.5]{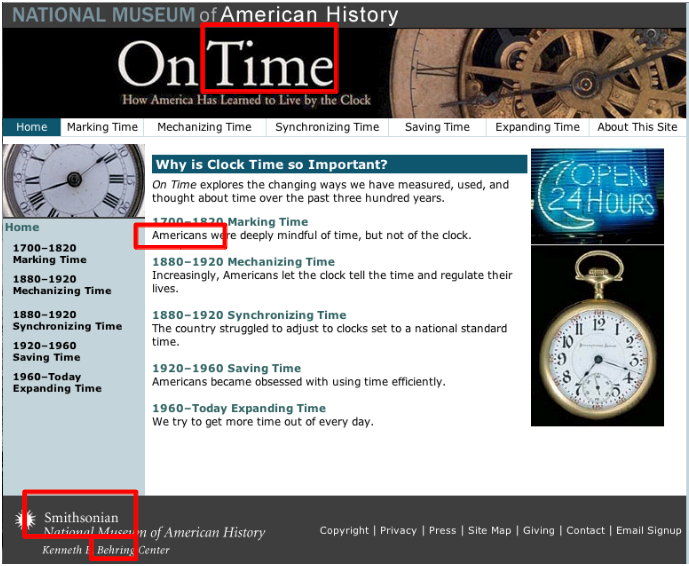}
 \caption{Example Page with terms of Lexical Signature highlighted}
 \label{fig:lex-sig}
\end{figure}
A lexical signature, as used previously and in this experiment, is a small number of terms chosen to represent the ``aboutness'' of a webpage.
Various methods have been proposed for calculating lexical signatures, but each attempts to choose terms based on the commonality of the term on the page, and/or the rareness of that term in the corpus.
Figure \ref{fig:lex-sig} shows an example of a lexical signature for a webpage, with the terms of the lexical signature outlined in red.
The words ``Smithsonian'' and ``Behring'' have been chosen those words are rarely used in the corpus, so their presence here might indicate something important about the nature of this page.
The words ``Time'' and ``Americans'', though common in the corpus, are nevertheless used repeatedly on this page, indicating that they are important to the meaning of this page.
Indeed, since the page is about the concept of time in America, and hosted by the Behring Center, part of the Smithsonian, this lexical signature provides a reasonably unique, if terse, description of the page.

The underlying principle, which has been demonstrated in prior studies, is that this lexical signature can be used to find or identify the page that it represents.  
Our experiment is focused on calculating a lexical signature for a URI when no representation of that URI can be retrieved.

\subsection{TFIDF}
To calculate lexical signatures, the Term Frequency-Inverse Document Frequency method was used to rank all available terms.
\begin{equation}
  TF_{i,j} = n_{i,j}
  \label{eqn:tf}
\end{equation}
Term Frequency is calculated for each term for each target URI, as shown in Equation \ref{eqn:tf}.
The TF of term \emph{i} for target URI \emph{j} is the number of times that \emph{i} appears in the backlinks of \emph{j}.
\begin{equation}
  IDF_{i} = log~\frac{|D|}{\left \{ d : t_{i} \in d \right \}}
  \label{eqn:idf}
\end{equation}
As shown in Equation \ref{eqn:idf}, Inverse Document Frequency (IDF) is the total number of documents in the corpus (\emph{D})
divided by the number of documents \emph{d} in which term \emph{i} occurs.
The Yahoo! API was queried to determine the frequency of that term on the web.
Klein and Nelson have shown \cite{1458510,klein:tc-df-correlation} that search engines can be used to approximate DF
for the purpose of lexical signature calculation.
\emph{$|D|$} was estimated using the numbers for the Yahoo! index taken from \newline
\url{http://www.worldwidewebsize.com/} on 11/1/2010.
\begin{equation}
  TFIDF_{i,j}= TF_{i,j} \times IDF_{i}
  \label{eqn:tfidf}
\end{equation}
TFIDF is the product of TF and IDF, as shown in Equation \ref{eqn:tfidf}, and the terms with the highest TFIDF values make up
the lexical signature.

In this way, TFIDF calculates the words that occur in the backlinks more often than they occur on the Web at large.
A lexical signature made up of such relatively rare terms should describe what the backlinks are about, and hopefully therefore describe what the target (missing) page is about.

Before TFIDF calculation, any terms included in a list of ``stop words'' were discarded.
A local cache of DF values was kept to reduce the number of queries to Yahoo!, on the assumption that DF values were unlikely to change significantly over the amount of time that lexical signatures were calculated, which was less than two weeks.

\section{Related Work} \label{sec:relwork}

Phelps and Wilensky \cite{phelps:robust} proposed calculating the lexical signature of a target page, and embedding that lexical signature into the link URIs to make the referenced page easier to find.
Their method relied on a five-term lexical signature being calculated at the time the link was created, and included in the link URI.
This placed the burden of preparing for future recovery on the content creator or administrator; if the creator did not calculate the lexical signature in advance, the user would be unable to use this method to attempt to rediscover the page.
In addition, web browsers would have to be modified to use the lexical signature in the URI to attempt to rediscover the page.
This meant that both web servers and browsers would have to implement this method for it to be usable.

Park, Pennock, Giles, and Krovetz \cite{1028101} expanded on the work of Phelps and Wilensky by analyzing eight different formulas for calculating a lexical signature.
Phelps and Wilensky's method involved dividing the number of times a term appeared in a document, known as its term frequency (TF), by the number of times that term appeared in all documents in the corpus, known as the document frequency (DF).
This fraction is known as term frequency-inverse document frequency (TFIDF).
Park et al. test Phelps and Wilensky's original TFIDF variant, a simpler TFIDF, plain TF, and plain DF as so-called ``basic LSs''.
In addition they tested several ``hybrid'' formulas in which some of the LS terms were calculated with one formula, and some of the terms with another formula, as potential ways to find relevant documents when the original document could not be found.
They found that TFIDF was the best among the ``basic'' formulas, though some of the hybrid formulas performed better in certain use cases.
They also used a five-term LS for lack of an empirical study on LS size, and noted that the effect of LS size was a topic for future research.

To remove the browser requirement in Phelps and Wilensky's system, Harrison and Nelson \cite{opal:ht06} proposed a system called Opal, wherein a webserver administrator would embed a JavaScript on HTTP 404 error pages to redirect users to an Opal server.
The Opal server would offer the user to view a cached copy of the page (if one could be found), or looking for another copy of the page using a lexical signature calculated from a cached copy.
This had two major advantages over Phelps and Wilensky's system.
First, a server administrator would have to make only one change, replacing the 404 error page, one time.
Second, this method required no modification to web browsers.
However, this still required some interaction on the part of the server administrator, and if the administrator did not implement this system, or if the server was unreachable, this system would not be usable.

Klein and Nelson \cite{klein:revisiting} proposed a method to use LSs calculated from archived or cached versions of a page to calculate a lexical signature.
They showed that a five-term or a seven-term lexical signature would produce the best results.
A seven-term LS did best at finding the URI in the first result, whereas a five-term LS performed best at finding the URI somewhere in the first page.

Henzinger, Chang, Milch, and Brin \cite{henzinger:news} used lexical signatures derived from newscast transcripts to find articles related to the newscast in real-time.  
Their input, rather than being a static web page, was a constantly-flowing stream of text from the transcript.
Their method took into account the temporal locality of terms, that is words that were spoken close together in the broadcast, to attempt to compute LSs that would be relevant to a single story each, rather than spanning across subsequent stories.
Their observations showed that, far from the five-term LSs used in prior studies, a two-term lexical signature worked best in this application.

Craswell et al. \cite{Craswell:2001:ESF:383952.383999} showed the effectiveness of anchor text in describing a resource.
They demonstrated that for a specific user need, which they called the site-finding problem, anchor text of backlinks provided a more effective way of ranking documents than did the content of the target page itself.

Sugiyama, Hatano, Yoshikawa, and Uemura \cite{sugiyama:refinement} proposed enhancing the feature vector of a web page by including its link neighborhood.
That is, they proposed that a search engine could more accurately describe the contents of a page by including information from both in-links (backlinks) and out-links.
They tested up to third-level in- and out-links, and found that only links up to the second level were helpful.
In some of the methods they tested, only the first-level links were helpful.

\begin{figure}[t!]
 \centering
 \includegraphics[scale=0.6]{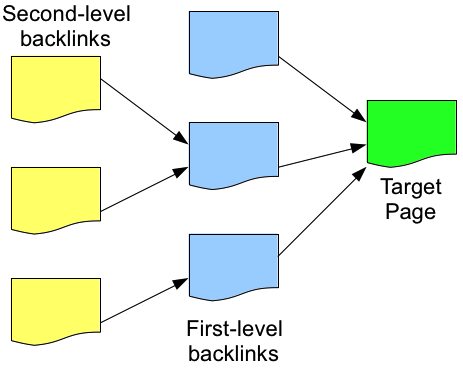}
 \caption{Example Link Neighborhood}
 \label{fig:link-neighborhood}
\end{figure}
Fujii et al. \cite{fujii:exploiting} explored the correlation between anchor text and page titles.
They showed that the collective nature of anchor texts, since they are created by many people, adds a significant value.
Anchor texts are created by a similar thought process to queries, and as such will use similar words to describe a topic.
Since links can be made by many authors, they will use their own word preferences, which means that anchors can provide synonyms that the original page's author might not use.
They even showed that anchor texts, since they are made by different authors and might be written in different languages, might be used to provide a bilingual corpus for machine learning of natural language translation.
\section{Methodology} \label{sec:method}
For the experiment, 309 URIs were obtained from the same corpus used
in Klein and Nelson 2008 \cite{klein:revisiting}.  For each URI, the
Yahoo! BOSS\footnote{\url{http://developer.yahoo.com/search/boss/}}
Application Programming Interface (API) was queried to determine the pages
that link to the URI (``backlinks'').  The Yahoo! BOSS API was used because
it was previously shown to give more complete backlink results than 
other search engines \cite{1255237}.  The order in which these backlinks
were returned is referred to as ``backlink rank''.  The backlinks of the
backlinks were obtained to create a directed graph of depth 2.  Figure
\ref{fig:link-neighborhood} shows an example of such a link neighborhood.
The page on the right, in green, represents the target page, the page that
a user linked to but is no longer available.  Using Yahoo!, we obtain
three pages that link to the target page.  These are the first-level
backlinks, represented in blue.  For each first-level backlink, we obtain
its backlinks, represented in yellow as the second-level backlinks.

%
%
\begin{table}
\centering
\begin{tabular}{l || r} \hline
Input URIs			&	309		\\ \hline
Runs per URI			& 	240		\\ 
~~Radius Combinations	&	5		\\
~~Backlink Depth Options	&	2		\\
~~Backlink Rank Options	&	3		\\
~~LS lengths			&	8		\\ \hline
Total URIs retrieved		& 	335,334	\\
Yahoo!  BOSS queries	&	154,858	\\
DF calculations			&	1,310,551	\\
Storage Used			& 	97 GB	\\
MySQL db size			&	5 GB		\\
\hline
\end{tabular}
\caption{Various Measures of Experimental Cost}
\label{tab:exp-costs}
\end{table}
Table \ref{tab:exp-costs} shows the various parameters used in constructing the 
test corpus.
It should be noted that many of these values are only so dramatic because we attempted some very costly options that turned out not to be helpful.
For example, we had to calculate DF for every term found on every page, including both first and second-level backlinks.
In practice, it would be possible to drastically reduce these costs, as DF would only have to be calculated for the terms used in anchor texts.
Additionally, one could avoid the second-level backlinks altogether, reducing the total number of backlink queries and page downloads by around 90\%.
All of the URIs thus obtained (the target URIs and the first and second-level backlinks) were retrieved and the representations saved to disk.
Five attempts were made to download any URI that did not result in HTTP success (200-level) codes.
\begin{table}
\centering
\begin{tabular}{l | r} \hline
\textbf{Level}&\textbf{\# of URIs}\\ \hline \hline
0	&	309		\\
1	&	28325	\\
2	&	306700	\\ 	\hline
Total	&	335334	\\
\hline
\end{tabular}
\caption{URI Counts at Backlink Depth Levels}
\label{tab:counts-levels}
\end{table}
Table \ref{tab:counts-levels} shows the total number of URIs discovered in this manner, at each level.  Level 0 are the URIs that started the experiment, level 1 are the URIs that link to level 0, and level 2 are the URIs that link to level 1.
\subsection{Pre-processing Link Neighborhood}
In order to guard against noisy inputs, five filters were applied to the backlink page representations before lexical signatures were calculated.  
Those filters were based on: content language, file type, file size, HTTP result code, and the presence of ``soft 404s''.
\subsubsection{URI canonicalization}
Since two URIs can have subtle textual differences while pointing to the same resource, it is necessary to canonicalize URIs\footnotemark.
For each URI we used, we downcased the domain, that is the section after the scheme identifier but before the next slash.
We removed any reference to port 80 in the domain portion of the URI, since 80 is the canonical port for HTTP and is therefore unnecessary.
We removed fragment identifiers, indicated by a hash sign (\#) and anything following it.
Finally, we removed a slash if it was the last character in the URI.
\footnotetext{The PHP function that we wrote to canonicalize URIs can be found at \url{https://github.com/jebstuart/researchtools}.}
\subsubsection{Filetypes}
Finding links in non-HTML documents is made difficult by the hundreds of potential document formats, so non-HTML representations were discarded.  In cases where the server described the file type in the HTTP headers, this information was trusted.  In other cases, the Unix `file' command-line program was used to guess the filetype.
Table \ref{tab:filetypes} shows the percent of pages characterized as various filetypes in this way.  
Fortunately, most pages retrieved were HTML.

\begin{table}
\centering
\begin{tabular}{l || r} \hline
\textbf{Filetype}&\textbf{percent of pages}\\ \hline \hline
HTML			&	97.52\% 	\\
PDF				&	1.08\% 	\\
Code			&	0.90\% 	\\
MS Word			&	0.29\% 	\\
MS PowerPoint		&	0.11\%	\\
RSS				&	0.04\%	\\
Excel			&	0.02\%	\\
GZip				&	0.02\%	\\
Other Binary		&	0.02\%	\\
Atom				&	0.01\%	\\
RDF				&	0.01\%	\\
\hline
\end{tabular}
\caption{Filetypes}
\label{tab:filetypes}
\end{table}
\subsubsection{Backlink validation}
For each backlink we retrieved, the location in the page of the link pointing to the target URI was important.
This link would be used as the centerpoint around which we would draw words to create lexical signatures.
Since it was returned in a query for pages that link to the target URI, each backlink should have at least one link pointing to its target URI.
As such, we attempted to find and mark all of the links that should have existed using a script that checked all HTML Anchor tags (``$<$a$>$'').

\begin{figure}[t!]
 \centering
 \includegraphics[scale=0.3]{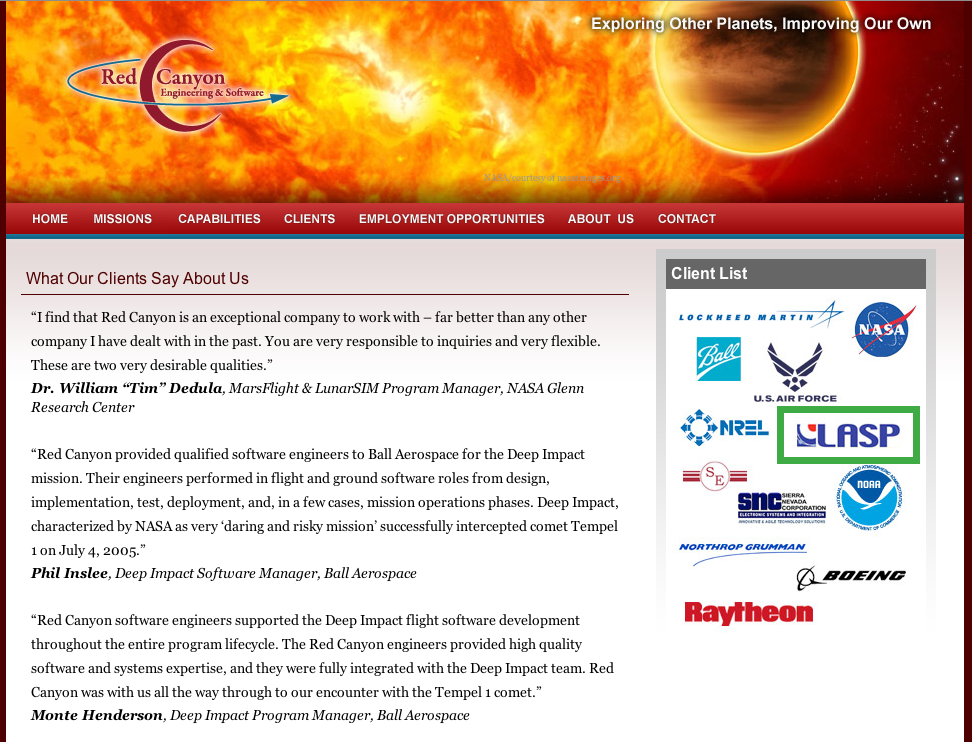}
 \caption{Example of an Image Map}
 \label{fig:image-map}
\end{figure}

In some cases, the links were not found.  
For example, in some pages the link would only have been brought in by a JavaScript that loaded in a portion of the page dynamically; we did not execute any scripts.
In another exception, links can be contained in image maps.
An image map is an image on a web page in which different portions of the image can link to different URIs.
For example, in Figure \ref{fig:image-map}, the link with which we are concerned exists in the image map on the right of the page, and is highlighted in green.
However, there being no text in or around the link, this link is not useful as a centerpoint for choosing words for a lexical signature.
For this reason, we decided not to use pages where the only link to the target URI existed in an image map.

In order to maintain knowledge of where the links existed in the page after rendering (see below for a description of rendering), a block of text consisting of the 32-character MD5 hash of the target URI was inserted before and after the link.
Searching for this block, which was reasonably assured to be unique within the page, yields the location of the link in the page after all HTML tags (including the tags which make the link) were removed.
These MD5 blocks were not included in lexical signature calculation.
\subsubsection{Rendering}
Lexical signatures are calculated only from the words that a human viewer of a page would see.
For this reason, we rendered the retrieved pages to remove HTML tags, style sheets, scripts, etc.
We used `lynx' for this purpose, using the following command:
\begin{quote}
lynx -nolist -force\_html -dump [original\_file] $>$ [rendered\_file]
\end{quote}
\subsubsection{Language}
Since an American search engine was being used, which is likely to have coverage biased toward English-language pages \cite{Vaughan2004693}, an attempt was made to discard non-English pages before lexical signature calculation.
The pages were analyzed with the Lingua Perl module\footnote{\url{http://search.cpan.org/~ambs/Lingua-Identify-0.29/README}} to obtain percentage guesses for each language in Lingua's standard dictionary.  

Table \ref{tab:languages} shows the percent of pages for which Lingua's top guess was the language shown.
Languages that accounted for less than 0.2 per cent of the results are not shown.
``English*'' represents those pages for which English was not the top guess, but the likelihood that the page was English was within forty per cent of the likelihood given to the top guess language.
These pages, as well as those for which English was the most likely language, were kept.
All others were discarded.
\begin{table}
\centering
\begin{tabular}{l || r} \hline
\textbf{Language}&\textbf{percent of pages}\\ \hline \hline
English			&	77.33\% 	\\
English*			&	14.80\% 	\\
Bulgarian			&	1.97\% 	\\
German			&	1.59\% 	\\
French			&	1.40\%	\\
Italian			&	0.82\%	\\
Polish			&	0.55\%	\\
Dutch			&	0.31\%	\\
Russian			&	0.24\%	\\
Turkish			&	0.24\%	\\
Spanish			&	0.21\%	\\
\hline
\end{tabular}
\caption{Languages}
\label{tab:languages}
\end{table}

\subsubsection{File size}
We discarded any pages that contained less than fifty tokens after rendering.  
This is similar to the filter applied by Park et al. \cite{1028101} and several subsequent studies to ensure enough input for a robust LS.  
Similarly, Henzinger et al. \cite{henzinger:news} chose their time window of 15 seconds of broadcast to correspond to roughly 50 words of input for LS calculation.

To reduce the amount of processor time needed, excessively large files were not completely analyzed by the language filter.
For files larger than 50 KB, only the first 50 KB of the file were analyzed by the language filter.
We assumed here that the language of a file would be reasonably consistent from beginning to end.

\begin{figure}[h!]
 \centering
 \includegraphics[scale=0.6]{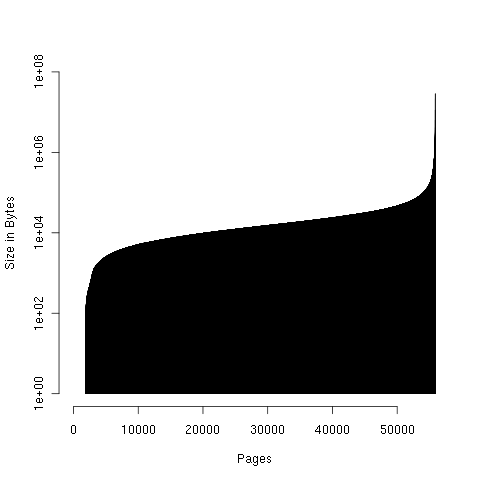}
 \caption{File Sizes}
 \label{fig:file-sizes}
\end{figure}
Figure \ref{fig:file-sizes} shows the file sizes of all retrieved URIs.
Each vertical line (which are so small they appear to be part of a block) represents one page, and its height is shown in bytes on a logarithmic scale.
Note that a few pages had zero-length results, shown by the white space on the left of the figure.
The majority of pages were between ten and one hundred kilobytes in length, and a few files reached into the hundreds of megabytes.  The largest was 129.8 MB.
\subsubsection{HTTP Status Codes}
Table \ref{tab:http_status_codes} summarizes in decreasing order of frequency the errors we experienced while
downloading all web pages.
Fortunately the vast majority of pages returned the HTTP response code $200$ which means success. We made a
total of five attempts to download pages that returned any other response code as well as pages causing any 
other of the liseted errors. If none of the attempts resulted in success, those pages were dismissed.
\begin{table}[h!]
\centering
\begin{tabular}{l || l || r} \hline
\multirow{0}{*}{\textbf{HTTP}}&\textbf{Status}&\textbf{percent} \\
\textbf{Response}&&\textbf{of pages} \\
\textbf{Code}&& \\ \hline \hline
200     &OK&                    96.75\%  \\
404     &Not Found&             1.27\%  \\
403     &Forbidden&             0.84\%  \\
        &Timeout&               0.42\%  \\
        &DNS Error&             0.14\%  \\
        &SSL Error&             0.14\%  \\
500     &Internal Server Error& 0.09\%  \\
        &No Output&             0.07\%  \\
        &Connection Refused&    0.06\%  \\
\hline
\end{tabular}
\caption{HTTP Status Codes}
\label{tab:http_status_codes}
\end{table}
\subsubsection{Soft 404s}
\begin{figure}[h!]
 \centering
 \includegraphics[scale=0.4]{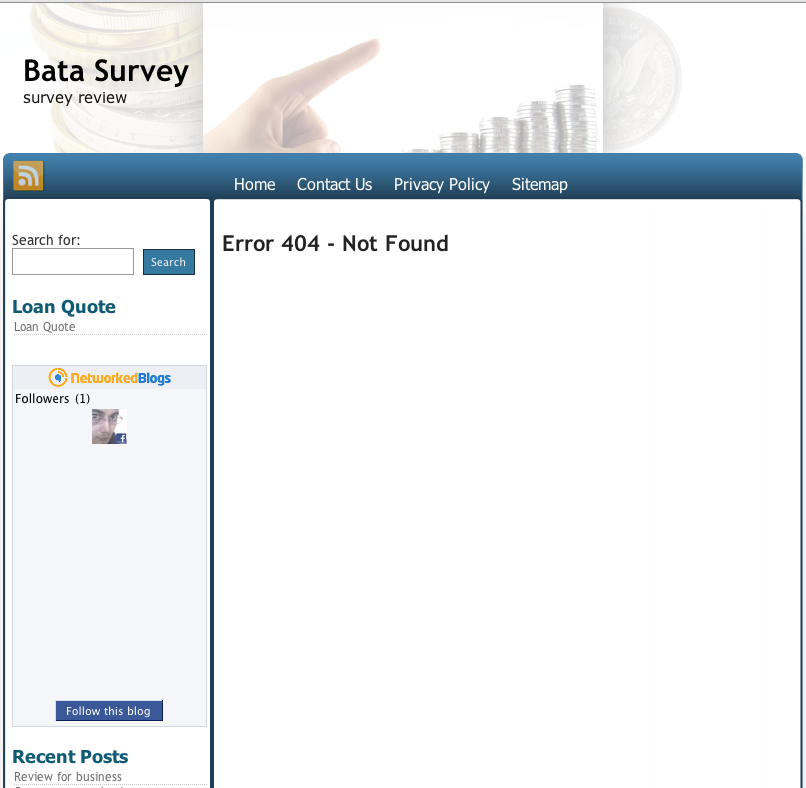}
 \caption{An example of a Soft 404 page}
 \label{fig:soft404}
\end{figure}

\begin{figure}[h!]
 \centering
 \includegraphics[scale=0.7]{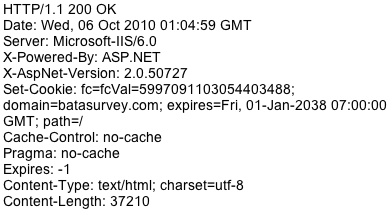}
 \caption{Headers for a Soft 404 page}
 \label{fig:soft404-headers}
\end{figure}

Pages that terminate with an HTTP 200 status code (possibly after
multiple redirects) while returning an error message in the human-readable
version of the page like ``your page could not be found'' are known
as ``soft 404s'' (for example, Figure \ref{fig:soft404}).  While to a
human user it is clearly a page not found, the headers shown in Figure
\ref{fig:soft404-headers} tell the browser software that the result code
is 200, meaning that the page {\bf was} found.


Since these pages are not ``about'' the same things as the pages that they link to, they were discarded.
To determine if a page was a soft 404, we used a subset of the method laid out by Bar-Yossef, Broder, Kumar, and Tomkins \cite{yossef:sic-transit}.

We retrieved the URI, plus two modified versions of the URI with a random 20-character string inserted.
First, we inserted the random string immediately after the first slash after the domain name.
Second, we inserted the random string immediately after the last slash in the path.
For example, if the original URI was
\begin{quote}
\url{http://example.com/path/to/file.html}
\end{quote}
and the random string was XXXXXXXXXXXXXXXXXX, then we would have downloaded also
\begin{quote}
\url{http://example.com/path/to/XXXXXXXXXXXXXXXXXXfile.html}
\end{quote}
and
\begin{quote}
\url{http://example.com/XXXXXXXXXXXXXXXXXXpath/to/file.html}
\end{quote}

The similarity of the representation returned for the original URI and both modified URIs was determined using the Jaccard similarity coefficient\footnotemark.
\footnotetext{The PHP function that we wrote to calculate the Jaccard similarity coefficient can be found at \url{https://github.com/jebstuart/researchtools}.}
\begin{equation}
  J(A,B)=\frac{|A\cap B|}{|A\cup B|}
  \label{eqn:jaccard}
\end{equation}
As shown in Equation \ref{eqn:jaccard}, the Jaccard coefficient of two pages A and B is the number of unique words that A and B share in common (their intersection) divided by the total number of unique words in A and B combined (their union).

If the original URI returned an HTTP error code, meaning anything other than a 200-level success code, then we could declare that the page was not a soft 404.
If the first modified URI returned a HTTP error code, then the page could not be a soft 404, since malformed URIs in that path trigger real (hard) 404 responses.
Otherwise, the Jaccard similarity between the representation of the original URI and the first modified URI, and the Jaccard similarity between the original URI and second modified URI were calculated.
If both similarity coefficients were above 0.9, then the original URI was deemed to be a soft 404 and discarded.

\begin{table}
\centering
\begin{tabular}{l || r} \hline
\textbf{Filter}&\textbf{percent of URIs discarded}\\ \hline \hline
Content Language	&	7.36\% 	\\
File Type			&	3.82\% 	\\
File Size			&	6.86\% 	\\
HTTP Result code	&	3.25\% 	\\
Soft 404s			&	0.36\%	\\	\hline
Remaining		&	78.35\%	\\
\hline
\end{tabular}
\caption{Pre-Processing Results}
\label{tab:pre-process}
\end{table}
Table \ref{tab:pre-process} shows the percent of URIs that were caught in each pre-processing filters.  
Some URIs were caught in more than one filter, but are only listed in the filter in which they were caught first.
\begin{figure}[t!]
 \centering
 \includegraphics[scale=0.5]{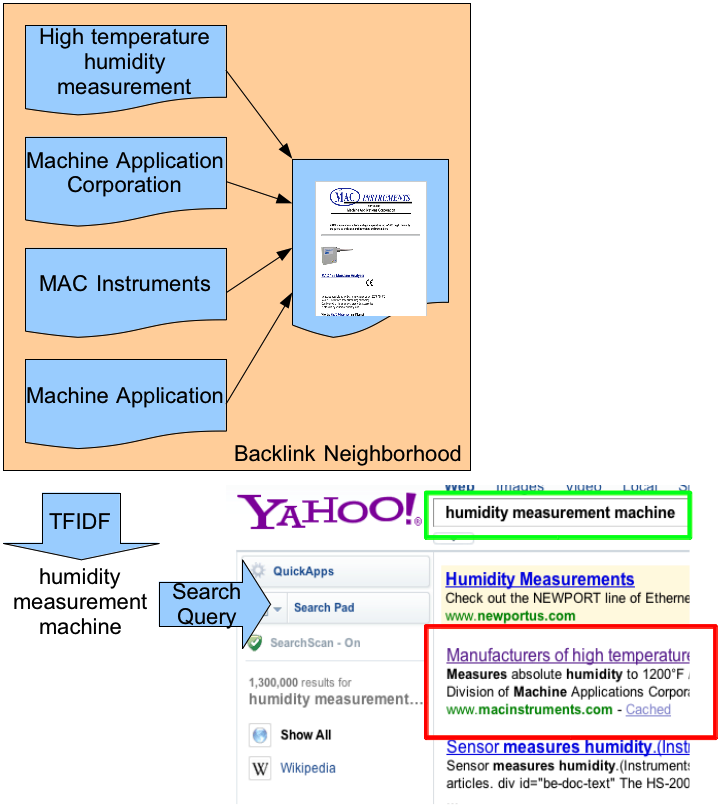}
 \caption{A 3-term lexical signature generated from the link neighborhood.}
 \label{fig:method-example}
\end{figure}
\subsection{Lexical Signature Calculation}
As mentioned in the Introduction, we sought to determine the effects of lexical signature size, backlink depth, backlink ranking, as well as the radius within a backlink page from which terms for the lexical signature should be drawn.  
For each of these factors, several possibilities were considered, and every possible combination of those possibilities was tested.
Using each combination, the TFIDF value of every term in the appopriate section(s) of the appropriate pages was calculated, and the terms with the highest TFIDF value were taken as the lexical signature. 

Figure \ref{fig:method-example} shows the example of the best-performing method applied to one URI.
First, we construct a backlink neighborhood of all first-level backlinks for the target URI, shown in the orange box.
Four backlink pages link to the target URI.
The anchor texts that they use to link are shown in the blue pages.
We take all of the words from the anchor texts, calculate TFIDF for each, and come up with a lexical signature consisting of the three words with highest TFIDF values.
We use that lexical signature as a query to the Yahoo! search engine, and search for the target URI in the results.
In this case, the target URI was the first result, which is the desired outcome.  
\subsubsection{Backlink Depth}
The two options for depth were (1) to use only the first-level backlinks, those that link directly to the target page and (2) to use first and second level backlinks. 

First-level backlinks being closer to the target page, they might result in a lexical signature that more accurately describes the missing page.  However in cases where few first-level backlinks exist, second-level backlinks might provide more information, leading to a better lexical signature.
\subsubsection{Backlink Ranking}
The backlinks returned from the Yahoo! BOSS API are ordered, starting at one and increasing for each result.  
To test whether this ranking was helpful, three possibilities were tested.  
First, only the top ten backlinks were used.  
Second, the top hundred backlinks, and third, the top thousand backlinks.  
If fewer backlinks existed than were allowed by the limit, as was typically the case in the top-thousand case, then there were fewer than one thousand backlinks used. 

If the rankings in backlink results from the BOSS API were helpful, then using only the top backlinks would provide a better lexical signature.  If the rankings of backlink results were not helpful, then using as many backlinks as possible might provide the better lexical signature by including more data.
\begin{figure}[t!]
 \centering
 \includegraphics[scale=0.4]{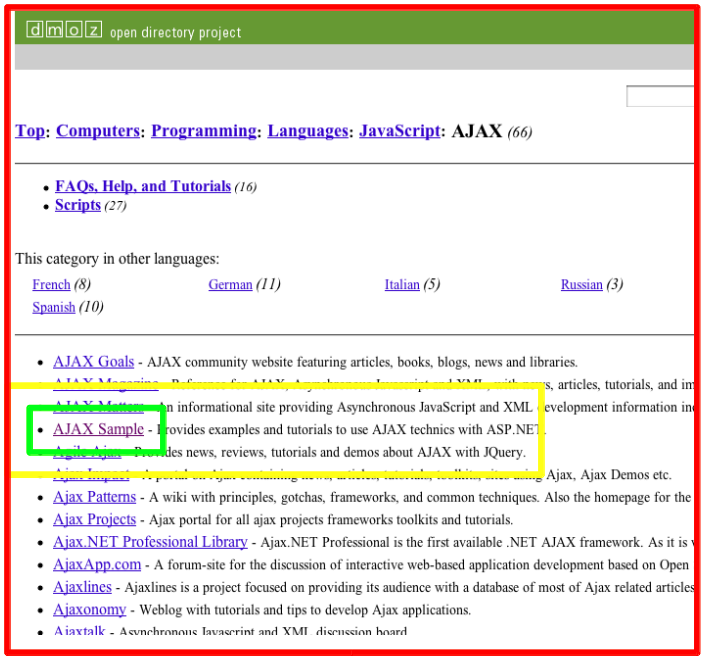}
 \caption{An example of Radius}
 \label{fig:radius-example}
\end{figure}
\subsubsection{Radius}
Four possibilities were considered for the radius within the backlink page from which lexical signatures would be drawn.
Figure \ref{fig:radius-example} shows the concepts of radius that were approximated.
Lexical signatures are typically drawn from entire pages, and this was used as the first possibility, shown in the red box.
However, since a particular section of a page can be about a different topic, or a more specific topic, than a page as a whole, it was tested whether using only the relevant portions of a page would produce a better lexical signature.
To find the ``relevant'' portion of a backlink page, the link from the page to the target URI was used as a centerpoint.  
Such a link should exist, since the backlink was described by the BOSS API as a page that linked to the target URI.

Using the link to the target page as a centerpoint, an attempt was made to capture the `paragraph' of context around the link, shown in yellow.
To do this, the second option considered the anchor text plus the five words before and five words after.
Option three included the link plus ten words on each side.
Finally, the fourth option used only the anchor text itself, shown in green.

In cases where a given backlink page included multiple links to the same target URI, the text around every link was included.  
In cases where the link to the target URI could not be found, the backlink page was not included in the calculations.
\subsubsection{Lexical Signature Size}
Most research using lexical signatures recommends using 5- or 7-term lexical signatures.
However, given that the lexical signatures in this experiment were being derived from a link neighborhood instead of the target page itself, the applicability of those standards was tested.  The ten terms with highest TFIDF value were stored, in order, and queries were performed using lexical signatures of sizes one, two, three, four, five, six, seven, and ten.
\subsection{Scoring of Results}
Since users tend not to look past the first few results for a 
query \cite{Agichtein:2006:IBB:1150402.1150526,Joachims:2005:AIC:1076034.1076063}, a scoring system that rewarded results at 
the top of the result set and gave less value to results past the first page was necessary.
For this purpose, normalized Discounted Cumulative Gain (nDCG) was used, with a relevance score of 1 for an exact match of the target URI, and 0 otherwise.
Discounted Cumulative Gain (DCG) is calculated using the relevance score for each item \emph{i} in an ordered result set of size \emph{p}, according to Equation \ref{eqn:DCG}.

\begin{equation}
  DCG = \sum_{i=1}^{p} \frac{2^{rel_{i}} - 1}{\log_{2} (1+i)}
  \label{eqn:DCG}
\end{equation}

nDCG is calculated by taking the Discounted Cumulative Gain and dividing by the Ideal Discounted Cumulative Gain (IDCG) produced by a perfect ordering of the result set.  
This formula for nDCG is show in Equation \ref{eqn:nDCG}.
\begin{equation}
  nDCG = \frac{DCG}{IDCG}
  \label{eqn:nDCG}
\end{equation}
Since we have binary relevance (i.e., one result has a value of 1 and all others have a value of 0), the IDCG will always have a value of exactly 1, so nDCG is equal to DCG.
Further, since only one member of the result set has a non-zero relevance score, and that one result always has a relevance score of exactly 1, we can simplify the DCG equation as shown in  Equation \ref{eqn:DCGstar}, where \emph{i} is the position of the target URI in the result set.
\begin{equation}
  DCG* =  \frac{1}{\log_{2} (1+i)}
  \label{eqn:DCGstar}
\end{equation}
Only the first 1000 members of the result set were checked.  If the target URI wasn't found, we assigned a nDCG value of 0, corresponding to an infinitely deep position in the result set.
\begin{figure}[h!]
 \centering
 \includegraphics[scale=0.6]{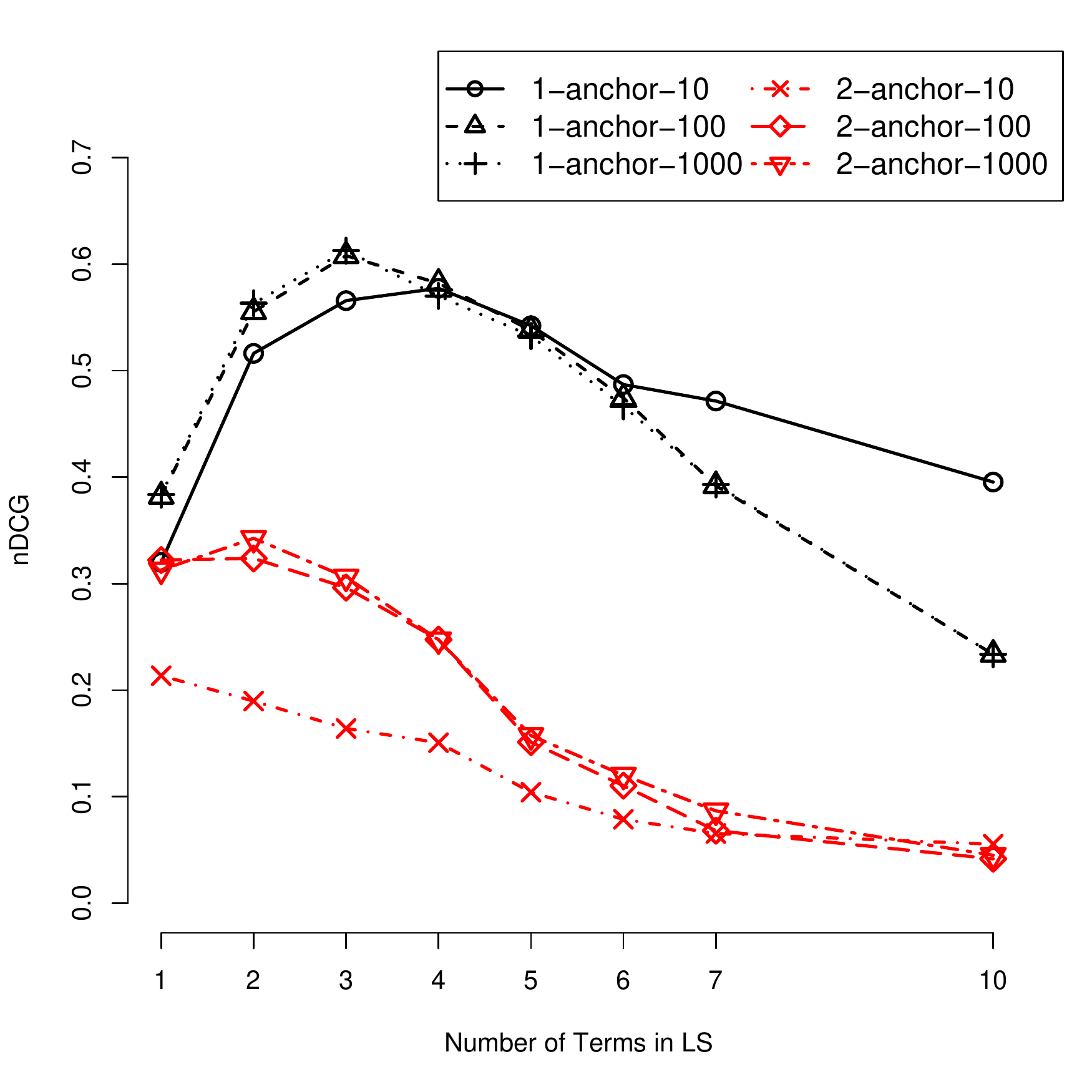}
 \caption{First- and Second-Level Backlinks \textbf{Anchor} Radius Lexical Signatures with Various Backlink Ranks (shown as levels-radius-ranks)}
 \label{fig:anchor-backlinks}
\end{figure}
\begin{figure}[h!]
 \centering
 \includegraphics[scale=0.6]{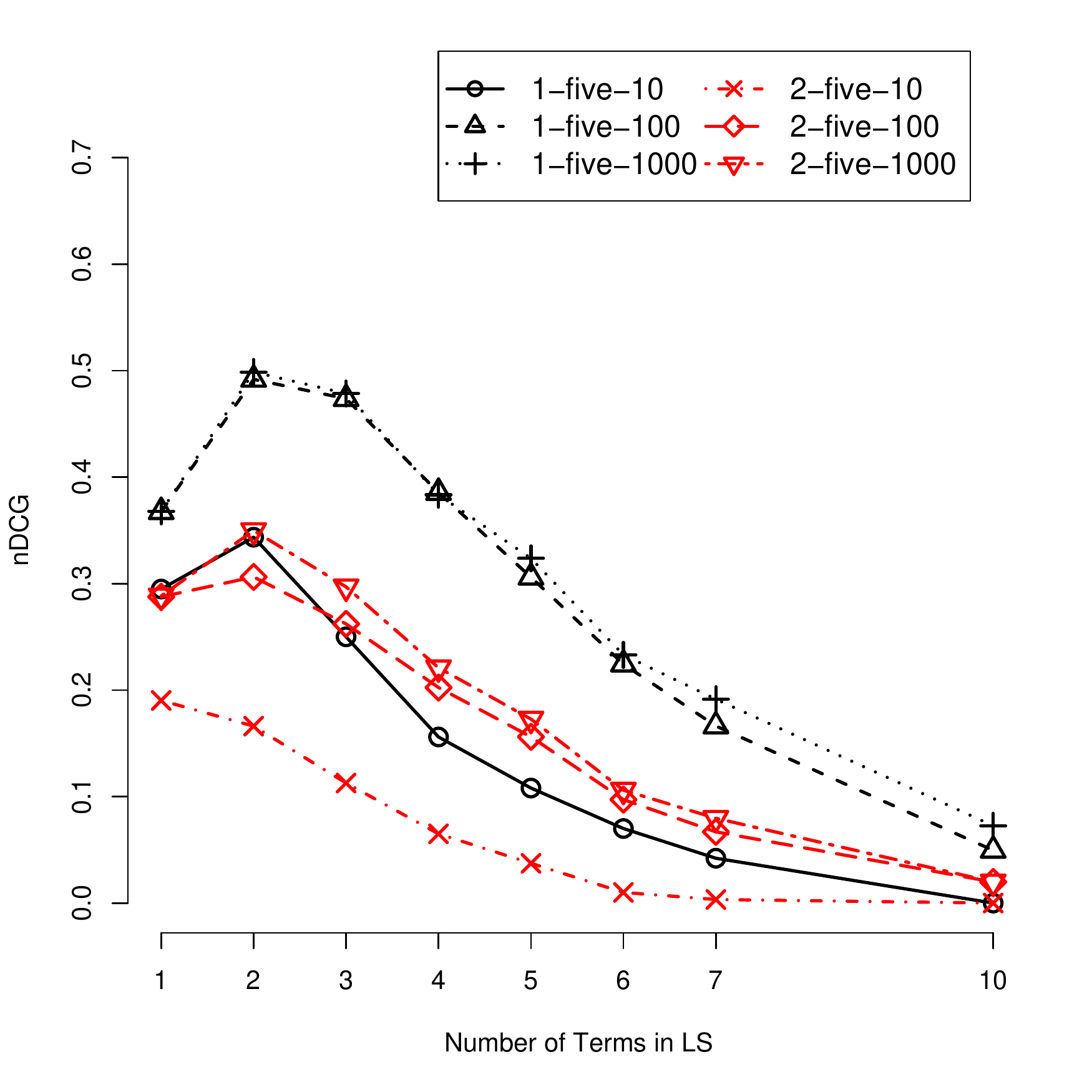}
 \caption{First- and Second-Level Backlinks Anchor \textbf{Plus/Minus Five} Radius Lexical Signatures with Various Backlink Ranks (shown as levels-radius-ranks)}
 \label{fig:five-backlinks}
\end{figure}
\begin{figure}[h!]
 \centering
 \includegraphics[scale=0.6]{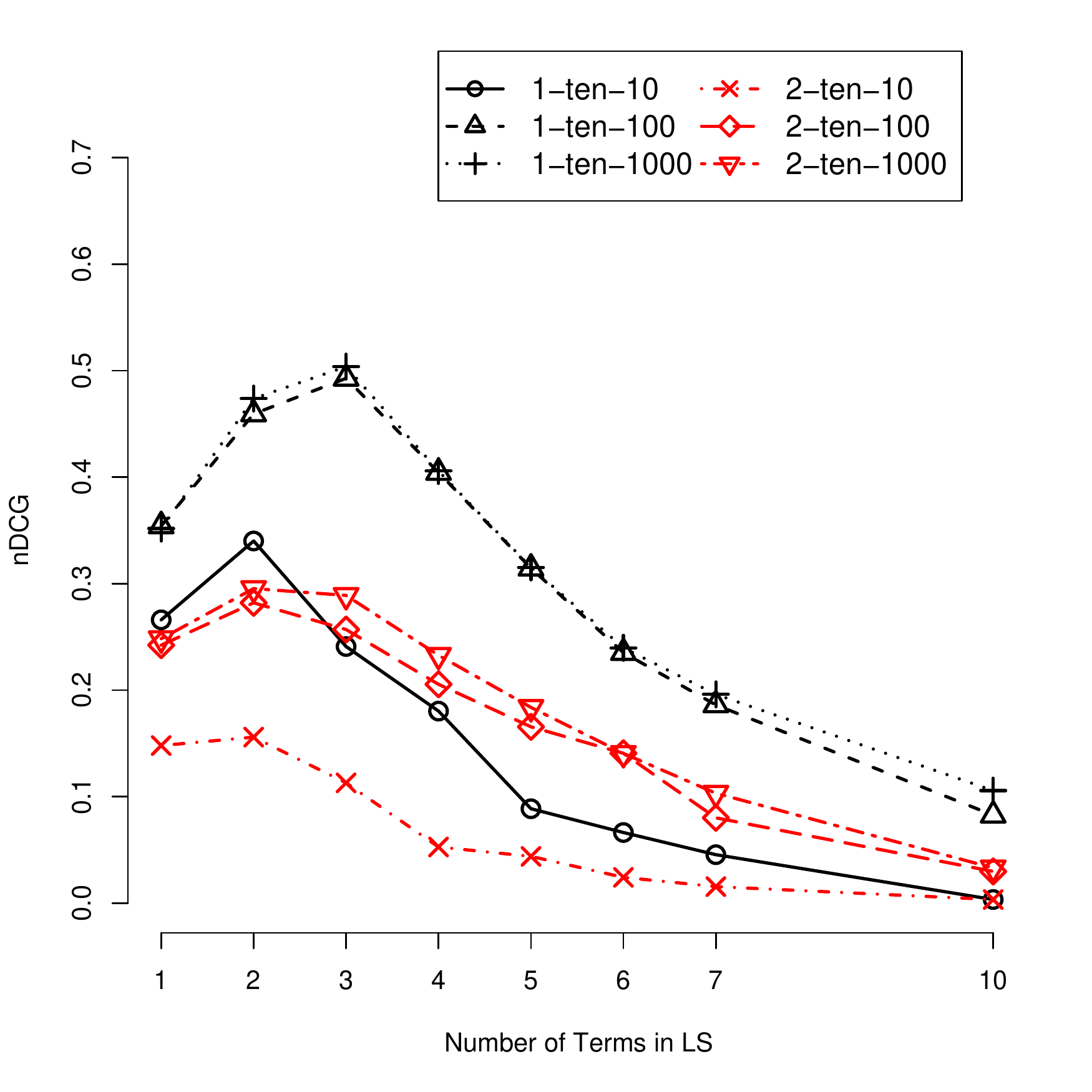}
 \caption{First- and Second-Level Backlinks Anchor \textbf{Plus/Minus Ten} Radius Lexical Signatures with Various Backlink Ranks (shown as levels-radius-ranks)}
 \label{fig:ten-backlinks}
\end{figure}
%
%
%
\section{Results and Discussion} \label{sec:results}
First, we discuss how we chose the best-performing method for calculating lexical signatures from backlinks, then we discuss the success of that method.  
Note that, in all graphs, each line represents one method for determining lexical signature.
The X-axis is the number of terms included in the lexical signature, and the Y-axis is the mean nDCG for that method.
\subsection{Backlink Depth}
%
Figures \ref{fig:anchor-backlinks}, \ref{fig:five-backlinks} and \ref{fig:ten-backlinks} show average
scores of methods based on anchor text, anchor text $\pm5$ words and anchor text $\pm10$ words, 
respectively, using the first- and second-level backlinks. First-level backlink methods are drawn in black
and second-level methods in red. Since the method based on using the whole
page performed execptionally poorly and its results do not change the overall outcome we decided to
not include the according graph.
Note the dramatic decline in every case when second-level backlinks are included.
This shows that second-level backlinks' relation to the target page is not tight enough to be useful
in describing the target page.
As such, our best-performing method includes only first-level backlinks.
\begin{figure}[t!]
 \centering
 \includegraphics[scale=0.6]{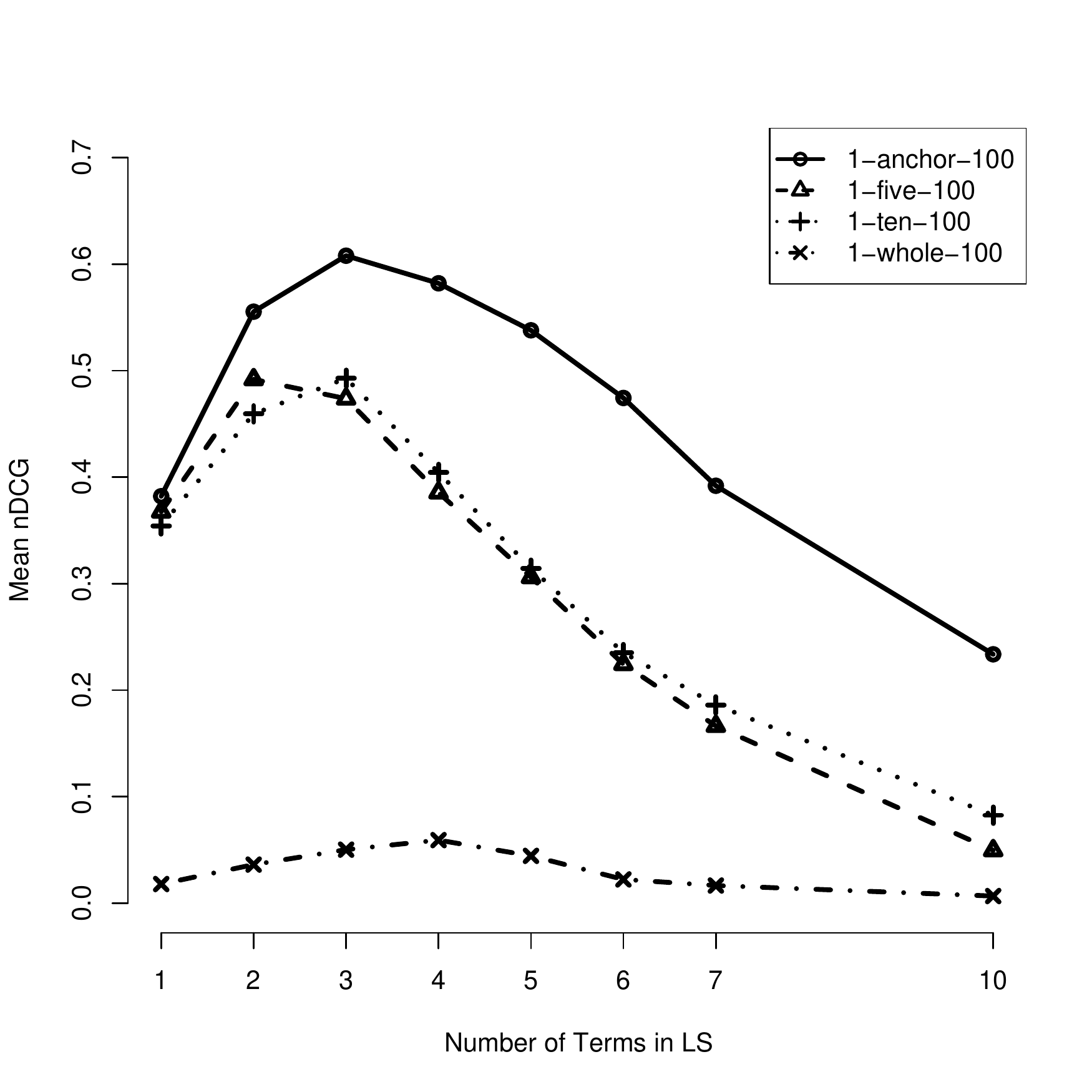}
 \caption{Effect of Radius (First-Level Backlinks)}
 \label{fig:radius}
\end{figure}
\subsection{Radius}
We started with the assumption that some parts of a backlink page would
use terms that are more closely related to the target page, and that
the most relevant terms would be in or near the link to the target URI.
In Figure \ref{fig:radius} we see that by far the best results arise from
using only the terms in the anchor text itself to calculate the lexical
signature.  The anchor text $\pm5$ words or $\pm10$ words performed
similar to each other, and using the whole page performed the worst.
Each step taken away from the anchor text, by broadening the radius to
include words around the anchor or the entire page, yields poorer and
poorer results.
\begin{figure}[t!]
 \centering
 \includegraphics[scale=0.6]{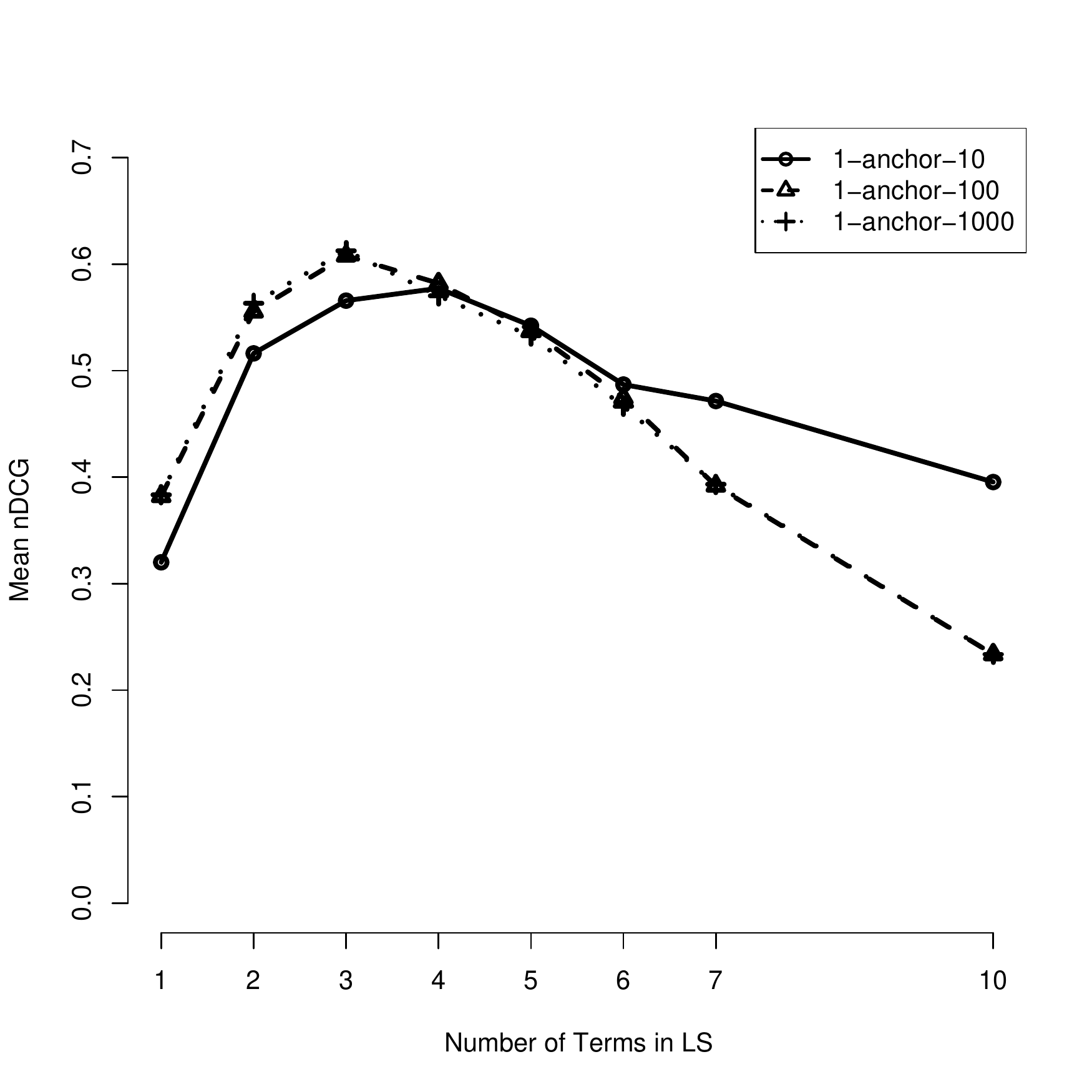}
 \caption{Effect of Backlink Rank (First-Level-Backlinks)}
 \label{fig:backlink-rank}
\end{figure}
\subsection{Backlink Ranking}
%
Figure \ref{fig:backlink-rank} shows the three possibilities for backlink ranks:
using only the top 10, top 100, or top 1000 backlink results.  
Note that using $100$ and $1000$ backlinks results in the highest overall nDCG value of $0.61$
obtained with 3-term lexical signatures. The corresponding nDCG value using only the top ten 
backlinks is $0.57$. 
However, considering the implied cost (see Table \ref{tab:exp-costs}) to acquire ten or
one hundred times as many pages and generate a lexical signature based on an accordingly larger
bucket of words, we consider using only the top ten backlinks as the more reasonable option. 
The ``return on investment'' is better while only sacrificing a drop of $0.04$ in nDCG.
So while we do not know how the ranking is determined by Yahoo!, we do know (consindering all costs)
that we are better off using only the top ten backlinks of a URI.
\subsection{Lexical Signature Size}
Also in Figure \ref{fig:backlink-rank} we can see that the overall best performance is obtained using
3-term lexical signatures. However, given the reasoning from the previous section, we consider the
best-performing lexical signature to be four terms in length.  
Using 4-term lexical signatures based on the top ten backlinks yields an nDCG of $0.58$.
Using more terms yields poorer results. Using fewer terms and top ten backlinks results in a slightly
worse performance compared to top $100$ and top $1000$ backlinks.
This result is noteworthy in that most other implementations of lexical signatures found that the best lexical signature size is five or seven terms \cite{klein:revisiting}.
The reason for the disparity is the source of the terms that make up the lexical signature.
In this method, the terms are drawn not from the target page itself, but from pages that link to it, which are likely to be ``related''.
Using five or seven terms drawn from the backlink pages is likely to over-specify the backlink pages
and their specific focus, rather than the content of the target page.
By using fewer terms, we decrease the risk of including a term in the lexical signature that does
not appear in the target page.
\subsection{Measures of Success}
As was noted before, the different methods for calculating lexical signatures were compared to 
each other using mean nDCG. This score represents the value of the result set to the user.
To see the meaning of this score more clearly, Table \ref{tab:rank_vs_ls_size_1000} shows the percentage
of URIs by their location in the result set using the first level backlinks, anchor text only and
the top 1000 results. Table \ref{tab:rank_vs_ls_size_10} show the same results for using the top ten
backlinks only.
The first row contains the percent of URIs that were returned as the number one result in the result set.
Each of these URIs received the maximum score of 1.
The following four rows represent those URIs that were on the remainder of the front page (2-10), 
further pages, and those that were not rediscovered at all.
The last row shows mean nDCG scores, for comparison. 

Note again that the best results were obtained using 3-term lexical signatures and the top
1000 backlinks. Over 58\% of the missing URIs were found in the first spot of the results. 
However, the top performance of our preferred method of using only the top ten backlinks 
and 4-term lexical signatures accounted for almost $56\%$ URIs returned at the top position. 
Less than $5\%$ of the URIs were returned between rank one and 1000.
These results confirm the ``binary distribution'' introduced in previous 
work \cite{klein:revisiting} meaning the majority of the URIs are either returned top 
ranked or somewhere (if at all) beyond the top 1000. 
\begin{table}[ht!]
\centering
\begin{tabular}{l || r | r | r | r | r | r | r | r} \hline
~ &\multicolumn{8}{ c }{\# of terms in LS} \\ \hline
\textbf{Result Rank}&\textbf{1}&\textbf{2}&\textbf{3}&\textbf{4}&\textbf{5}&\textbf{6}&\textbf{7}&\textbf{10} \\ \hline \hline
1 		& 32.11	& 50.50	& {\bf 58.19} 	& 54.85	& 52.51 & 45.82	& 38.80 & 23.41	\\
2-10		& 10.03	& 10.70	& 7.02	& 5.35	& 2.34	& 2.34	& 1.67	& 0.33	\\
11-100		& 5.69	& 3.34	& 0.67	& 0.33	& 0.33	& 0.33	& 0.33	& 0.67	\\
101-1000	& 4.35	& 1.67	& 0.00	& 0.00	& 0.33	& 0.33	& 0.33	& 0.00	\\
$>$ 1000	& 49.16	& 35.12	& 35.45	& 40.80	& 45.82	& 52.51	& 60.20	& 76.92	\\ \hline
Mean nDCG	& 0.38	& 0.56	& {\bf 0.61}	& 0.57	& 0.53	& 0.47	& 0.39	& 0.23	\\
\hline
\end{tabular}
\caption{Result Rank and nDCG vs Lexical Signature Size (1-anchor-1000)}
\label{tab:rank_vs_ls_size_1000}
\end{table}
\begin{table}[ht!]
\centering
\begin{tabular}{l || r | r | r | r | r | r | r | r} \hline
~ &\multicolumn{8}{ c }{\# of terms in LS} \\ \hline
\textbf{Result Rank}&\textbf{1}&\textbf{2}&\textbf{3}&\textbf{4}&\textbf{5}&\textbf{6}&\textbf{7}&\textbf{10} \\ \hline \hline
1 		& 25.08	& 45.15	& 52.51	& {\bf 55.85}	& 52.84 & 47.83	& 46.49 & 39.13	\\
2-10		& 9.03	& 9.70	& 7.02	& 3.34	& 2.01	& 1.34	& 1.00	& 0.67	\\
11-100		& 8.03	& 4.68	& 2.01	& 0.67	& 0.67	& 0.33	& 0.33	& 0.33	\\
101-1000	& 5.69	& 2.34	& 0.33	& 0.67	& 0.33	& 0.33	& 0.33	& 0.00	\\
$>$ 1000	& 52.17	& 38.13	& 38.13	& 39.46	& 44.15	& 50.17	& 51.84	& 59.87	\\ \hline
Mean nDCG	& 0.32	& 0.52	& 0.57	& {\bf 0.58}	& 0.54	& 0.49	& 0.47	& 0.40	\\
\hline
\end{tabular}
\caption{Result Rank and nDCG vs Lexical Signature Size (1-anchor-10)}
\label{tab:rank_vs_ls_size_10}
\end{table}
\section{Conclusions} \label{sec:conclusions}
In this paper, we showed that lexical signatures calculated from the backlink neighborhood of a web page can be used to re-discover that page.
This method can be used when no copies of the missing page exist, and the page was not analyzed before it went offline.

We found that only the first-level backlinks were related closely enough to the missing page to be helpful in creating a lexical signature.  
Including second-level backlinks introduced too much noise into the signatures.

The most relevant text in a backlink page is the text that is used to link to the missing page, and this ``anchor text'' provides enough terms to develop a viable lexical signature for the missing page.

The ranking of backlinks for a URI is meaningful and helpful in determining a lexical signature for the missing URI.  
We found that using only the first ten backlinks produced a better signature than including the first hundred or thousand backlinks.
The effect should be studied with greater granularity to determine the best number of backlinks to analyze, since we tested only values that were separated by an order of magnitude each.

The ideal size for a lexical signature derived from backlinks is three terms.  
Any fewer terms and the signature is too vague, likely to include the target URI deep in its results, if at all.
Any more terms and the signature runs the risk of including terms not present in the target page, which would likely exclude the target URI from the results.
\end{document}